\documentstyle[preprint,aps]{revtex}
\tightenlines

\begin{document}
\draft

\title{Scarring in vibrational modes of thin metal plates}

\author{Andre {\it J}. Starobin and Stephen W. Teitsworth} 

\address{Duke University, Department of Physics, Box 90305, Durham, NC 27708-0305}

\date{\today} 
\maketitle 

\begin{abstract} 
  We report the first direct experimental observation of scarring
  phenomenon  in transverse vibrational modes of a thin metal
  plate.  The plate has the shape of a full stadium and clamped boundary
  conditions.  
  Normal modes are imaged using time-averaged holographic
  interferometry, and modes corresponding to ``bouncing ball'' and higher
  order periodic trajectories are found. 
  An eikonal approximation of the
  solution along classical trajectories of the stadium including
  nontrivial phase shifts at clamped boundaries yields a useful quantization
  condition for the observed modes.
\end{abstract} 

\maketitle

Scarring refers to the build-up of amplitude along unstable periodic
trajectories in high order modes of certain wave systems. The term is
generally reserved for wave problems with domains which correspond to
classically chaotic billiards.  The phenomenon was first reported in
numerical studies of high order, short wavelength modes of the Helmholtz
equation inside a domain known as the Bunimovich stadium [1]; an
explanation of scarring relies on methods of quantum chaos [1,2].  The
Helmholtz equation serves, for example, as the time-independent
Schr\"{o}dinger equation for a particle in a box, and also as the equation
governing the transverse vibration of elastic membranes.  In addition to
numerical studies of the Helmholtz equation, several analogue experiments
have reported quantum chaotic spectral and spatial properties including
the electromagnetic field modes of quasi-two-dimensional microwave
cavities [3-5], stationary capillary waves on water [6], three-dimensional
acoustic resonances in water-filled cavities [7], and vibrational modes of
drumheads [8].

A possible extension of quantum chaos methods to the high frequency limit
of wave problems with modes that are not described by the Helmholtz
equation was noted by Berry [9].  The statistical properties of spectra as
predicted by random matrix theory have since been confirmed in experiments
on elastic waves in three-dimensional metal blocks [10-13], and on studies
of electromagnetic modes in three-dimensional microwave cavities [14,15].
In both of these cases the vectorial nature of the modes destroys the
exact analogy with the time-independent Schr\"{o}dinger equation.  More
recently, scar-like phenomena have been reported for the electromagnetic
modes of three-dimensional microwave cavities, although the role of
underlying periodic orbits is unclear [16].

The transverse vibrational waves of a thin plate provide another important
example of a wave system for which the stationary modes are not described
by the Helmholtz equation.  In a recent numerical study [17], the computed
asymptotic spectrum for a fully clamped thin plate was shown to possess 
statistical properties predicted by random matrix theory; furthermore,
the authors found strong evidence of scarring in vibrational amplitude
plots for some of the high frequency modes.  More recently, a general
theory of scarring and spectral properties has been developed by Bogomolny
and Hughes for transverse vibrations in thin plates [18].  In this paper,
we present experimental results on the high frequency properties of
vibrating metal plates, with particular attention given to the
distinguishing spatial properties of individual high order modes.  To our
knowledge, this is the first report of scarring phenomena in an
experimental vibrating plate.  We also present a quantization criterion
for the eigenfrequencies of modes that are scarred by particular classical
periodic orbits.  This criterion includes the effects of angle-dependent
phase shifts associated with reflection of periodic orbits at the fully
clamped boundary of a thin plate.

For our experiment, we have studied a stainless steel plate of thickness
$h = 0.305$ mm.  The plate is in the shape of a full stadium and consists
of a square central section of side $8.00$ cm and two semi-disks on each side
of radius $R = 4.00$ cm.  Contour plots of vibrational amplitude for
individual
modes are obtained using the technique of time-averaged holographic
interferometry [19].  Characteristic vibrational amplitudes are of the
same order as the wavelength of the 1 mW frequency-stabilized HeNe laser
used for imaging.  To achieve a fully clamped boundary condition (i.e.,
both amplitude and its normal derivative tend to zero at the boundary),
the plate edges were carefully epoxied to a massive aluminum support.  The
interferograms of all imaged modes show that the vibrational amplitudes
and amplitude gradients at the boundary are very small, thus confirming
the effectiveness of the clamping procedure.  To drive the plate 
vibrations harmonically, we use a modified audio speaker (frequency
range from 50 Hz to 12 kHz) which is coupled to the plate via a thin steel
rod which acts as the driving rod.  The rod is firmly attached to the
speaker voice coil at one end and lightly epoxied to the plate at the
other end.  A scannable phonograph cartridge stylus is used to monitor the
amplitude response of the plate at various positions.  The signal is
filtered through a low-noise pre-amplifier before being fed to an HP 3561A
signal analyzer which is used to measure the resonant frequency and
quality factor ($Q$) for each detectable mode.  Significant damping 
due to coupling of the plate to the support structure and the driving 
 rod sets in at approximately 12 kHz, at which point
the quality factors of the system become too low to resolve individual
modes.  The fundamental frequency of the plate is approximately 200 Hz.

In the frequency range between seven and eight kilohertz we find four 
vibrational resonances at $7272 Hz$, $7442 Hz$, $7622 Hz$ and at $7874 Hz$.
The wave patterns excited at $7442 Hz$ and at $7622 Hz$ reveal prominent scarring  
in the imaged amplitude distributions. Figure 1a shows the holographic
interferogram,  effectively an amplitude contour plot,
for the $7442$ Hz frequency of excitation.  The large, connected bright areas correspond to
places with
very small vibrational amplitude, while darker areas and their interiors
correspond to regions with large amplitude.  
Clearly visible in Fig. 1a 
is scarring by a ``bouncing ball" trajectory across
the width of the stadium plate; furthermore, we can see that there are 
7 antinode pairs as one traverses the complete periodic trajectory.
The excited wave pattern has a broken symmetry with respect to  
$\beta$ symmetry axis of the plate, with the symmetry axis of the plate 
and the position of the driving rod-to-plate contact designated in Figure 1b.
The symmetry with respect to $\alpha$ axis is partially preserved on the left side of the plate.
Further,
while the left side of the plate displays a clear build-up of amplitude
along the ``bouncing ball" orbit and has regular radial amplitude features
in left semi-disk, features characteristic of the (6,3) modes of a pure disk, 
the amplitude distribution in the right half of the plate does not appear regular 
and does not show any obvious scarring.

Figure 2 shows the interferogram
for the $7622$ Hz mode.  In this case, scarring on the 
right side of the plate clearly follows a rectangular orbit
 producing a ``whispering gallery" type mode.  The number of antinodes
traversed
in going around the right half of the orbit is 16, which suggests that the
total
number of antinode pairs for the full orbit is 16. This resonant wave pattern 
is almost perfectly localized on the right-hand side of the plate, the same side 
where the speaker rod is glued to the plate. The semi-rectangular
scar feature is confined fully to the right half of the plate and is 
essentially symmetric with respect to the $\alpha$ axis. Immediately to the left 
of the $\beta$ axis we find clustering of amplitude along a vertical located 
almost at the same position as the ``bouncing ball'' scar of the 7442Hz wave pattern.
As we move further to the left, the excitation amplitude decays, with 
the first bright fringe barely resolved. 

The broken symmetry of the excited wave patterns we observe clearly indicates
that they are not pure modes of the underlying unperturbed plate. 
In the 7-8 KHz range the experimental Q-factors, measured as the ratio of 
the resonance peak position to the 3dB width of the resonance peak, are between 50 and 90.
This corresponds approximately
to a minimum 75Hz level spacing that we can resolve experimentally. The observed 
resonance spacing in this frequency range is 200Hz which compares well with 
the mean Weyl level spacing for this plate of 300Hz. With such quality factors in the
experiment we can conclude that at each excitation frequency two-to-three pure modes
of the system are excited.

The 2nd Newton's law written for small amplitude flexural vibration of a
thin plate and the general form of the solution have the following form: 

	\begin{eqnarray}
	& -D\Delta^2\psi + F_{dephase} + F_{dr} = \rho\partial_{tt}\psi \\
	\psi({\bf r}, t; f) = & \sum_n \{ \alpha_n \cos(2\pi ft + \gamma)\phi_n({\bf r})
+ \beta_n\sin(2\pi ft + \gamma)\phi_n({\bf r}) \}    
	\end{eqnarray}
	$D$ is the flexural rigidity of the plate and $\rho$ is the mass
per unit area of the plate. $F_{dephase}$ and $F_{dr}$ are dephasing and driving 
forces respectively. $f$ is the excitation frequency and  $\gamma$ is a fixed 
time-independent phase shift.
The driving force in our experiment is harmonic and is localized at the 
position where the speaker needle is glued to the plate. In the absence 
of dephasing (with a proper choice of $\gamma$) $\alpha_n=\delta(f-f_n)$ 
and $\beta_n=0$. With a dephasing damping force both elastic and inelastic 
response is present and $\alpha_n, \beta_n$ are frequency dependent coefficients.
In this experiment damping occurs via plate-speaker and plate-air coupling, the 
former being the primary damping mechanism. Below, we write down expressions for 
imaged resonant wave patterns in terms of pure modes of the system and argue that these are also 
scarred.

The localization of wave pattern on the right at 7622 Hz excitation 
strongly suggests that we excite two closely spaced symmetry related modes: 
$\phi^{-+}_{i+1}$ and $\phi^{--}_{i+1}$, with the first (second) sign in the superscript
indicating odd (even) symmetry with respect to $\alpha$ ($\beta$) axis. The 
$\phi^{-+}_{i+1}$ and $\phi^{--}_{i+1}$  narrowly split mode pair 
located near 7622Hz and the $\phi_{i}$ mode slightly lower than 7442Hz are marked on the 
level diagram in figure 3. The opposite symmetry of the two $\phi_{i+1}$ modes
with respect to $\beta$ axis is 
needed to produce the observed localization. The odd symmetry with respect to 
the $\alpha$ axis of $\phi^{-+}_{i+1}$ and $\phi^{--}_{i+1}$
 is supported by the observed narrowness of the nodal 
line that runs along the $\alpha$ axis, clearly passing through the right semi-disk center
  (cf. fig. 2). For even symmetry one would expect 
a wider bright white region reflecting a wider interval of small excitation amplitudes. 
A much wider bright region is always seen near the clamped edges where the derivative 
vanishes (cf. fig 1 and 2) and the amplitude increases slower from zero than near the
$\alpha$ axis of the 7622Hz excitation pattern. The placement of $\phi^{-+}_{i+1}$ mode
on the level diagram higher than $\phi^{--}_{i+1}$ mode is motivated by a similar
consideration. The odd-even mode picks up extra strain energy compared to the odd-odd mode
along the symmetry line where its curvature is non-zero.

	With the level diagram of figure 3, the three mode expansion of the 
wave pattern excited at $7622Hz$ has the following 
form: 
	\begin{eqnarray*}
	\psi(7622) = \sin(2\pi f_{exc}t)\{\beta^{-+}_{i+1}\phi^{-+}_{i+1} +  
	\beta^{--}_{i+1}\phi^{--}_{i+1} + \beta_{i}\phi_{i}  \} +&	\\
	  \cos(2\pi f_{exc}t)\{\alpha_i\phi_i + \alpha^{-+}_{i+1}\phi^{-+}_{i+1} + 
	\alpha^{-+}_{i+1}\phi^{-+}_{i+1} \}
	\end{eqnarray*}
Since the two symmetry related modes are closely spaced with a spacing much smaller that the
mean level spacing, 
$\beta^{-+}_{i+1} \approx \beta^{--}_{i+1}$ and with 
 $\beta_{i} \ll \alpha_i \approx \alpha_{i+1} \ll \beta_{i+1}$ at 7622Hz, 
the excited wave pattern can be simplified to: 
	\begin{equation}
        \psi(7622) = \sin(2\pi f_{exc}t)\{\beta_{i+1}\phi^{R}_{i+1}\} + 
        \cos(2\pi f_{exc}t)\{\alpha_i\phi_i + \alpha_{i+1}\phi^{R}_{i+1} \}
        \end{equation}     

where $\phi^{R}_{i+1}$ is the sum(difference) of the two symmetry related modes with the 
sign chosen so to produce a wave pattern localized on the right side of the plate as is 
seen in the experiment. With the main contribution to this wave pattern localized on 
the right we expect to find the signature of the pure $ith$ mode of the plate on the left.
This we believe is the origin of the amplitude clustering along the vertical just off to the left 
of the $\beta$ symmetry axis.

	Similarly, we can now write the wave pattern excited at 7442Hz as :

	\begin{equation}
	 \psi(7442) = \sin(2\pi f_{exc}t)\{\beta_i\phi_i \} +
                  \cos(2\pi f_{exc}t)\{\alpha_{i+1}\phi^{R}_{i+1} + \alpha_{i}\phi_{i} \}
	\end{equation}

The perturbing term is fully localized on the right hand side of the plate.
We than expect to find a signature of the $ith$ mode of an unperturbed plate to the left
of the $\beta$ 
symmetry axis and a wave pattern arising  due to superposition of the $ith$ and 
 the $(i+1)st$ doublet on the right. Indeed, this is what we observe. The 
left side of the pattern in figure 1a is scarred by a ``bouncing ball'' orbit, while 
the right side exhibits no visible scarring, or regularity. 

	In an earlier work by Sridhar and Heller a localization effect similar to 
ours has been observed [9]. In that work resonant wave patterns of a MW cavity slightly 
perturbed from Sinai billiard shape were studied. 
There the authors of [9] were able 
to mix closely spaced symmetry related scarred modes of a Sinai shaped MW cavity, by slightly 
displacing the disk scatterer in the middle of the plate, with localization occurring on the 
side away from the displacement. When the symmetry of the cavity was reestablished the 
two mixed modes were split again. In our experiment we did not 
attempt to resolve individual symmetry related modes due to relatively low
Q-factors of the excitation system. 

To understand the scarring observed in high order modes within the simplest version 
of the PO formalism, we assume that
the time-independent part of the solution along the scarred trajectories has a
form $\displaystyle u = A_\Gamma e^{i S_\Gamma}$, where 
$A_\Gamma$ is
assumed to be a single valued function of position and is assumed to be
nearly constant in the neighborhood of a classical trajectory $\Gamma$[18].
$S_\Gamma(q_{\perp}, q_{\parallel})$ is a multivalued function written in
terms
of local coordinates $q_{\perp}$ and $q_{\parallel}$, along and
perpendicular to
the trajectory.

In the high frequency limit the form of $S_\Gamma(q_{\perp},
q_{\parallel})$ is
restricted. Far from the boundaries, sources, and caustics, $S_\Gamma$ is
nearly
linear in both $q_{\perp}$ and $q_{\parallel}$.  Moreover, $S_\Gamma$ is a
rapidly growing function of $q_{\perp}$ and $q_{\parallel}$ in the limit
of high order modes. In the regime where $S_\Gamma(q_{\perp}, q_{\parallel})$
is nearly linear, it is well described by an eikonal equation
for a thin plate:
\begin{equation}
       (\nabla S_\Gamma)^4 = \frac{12\rho(1-\sigma^2)}{Eh^2} {\omega_0}^2, 
\end{equation}
where $\omega_0$ denotes the frequency of the orbit.  Unlike the eikonal
equation for an ideal membrane, Eq. $(6)$ is
only valid up to frequencies for which volume deformations become
appreciable, i.e., such that $\displaystyle\frac{\Delta^2 u({\bf
r})}{\lambda^3}
\sim \frac{h^3}{\lambda^3} \sim 1$, where $u({\bf r})$ is the amplitude
function for the mode and $\lambda$ is the effective wavelength. 
At such short wavelengths and high frequencies the biharmonic wave equation
itself
breaks down.  For the experimental results reported here we never reach
this regime
because $\displaystyle\frac{h^3}{\lambda^3} \sim 10^{-3} \ll 1$.

 It can be seen that there are two types of solution of Eq. (6).  $S_\Gamma$ is
either a real or complex function of position.  Real $S_\Gamma$ gives rise to
propagating wave solutions which are plane wave-like in a homogeneous
plate, exactly as for a membrane. Complex $S_\Gamma$ yields exponentially
decaying, or rising solutions of the wave equation. Each mode of the plate
must be bounded, and hence the exponentially growing solutions cannot be
present far from the boundary. On the other hand, the decaying solutions
decay to zero for distances significantly larger than a wavelength. This
can be seen already in the one-dimensional clamped rod problem which
serves as a useful guide in understanding various aspects of the
two-dimensional 
plate [20,22].

It is straightforward to show that a plane wave propagating in a thin
plate incident at an angle $\alpha$ with respect to the normal onto a
clamped plate edge, is reflected with a phase shift given by:
\begin{equation}
        \delta =  -2\arccos\left(\sqrt{\frac{1 + 
                                 \sin^{2} \alpha}{2}}\,\right).
\end{equation}
This expression can be obtained in the standard way by matching incoming
and outgoing plane wave solutions under the condition that the wavelength
is much smaller than the local boundary curvature [18].  Then, a simple
quantization condition may be written for a trajectory $\Gamma$ with $N$
boundary reflections in the form:
\begin{equation}
    k_{\parallel}L - 2\sum\limits_{i=1}^{N}\arccos\left(\sqrt{\frac{1 + 
    \sin^{2} {\alpha}_{i}}{2}}\,\right) = 2\pi n,
\end{equation}
where $k_{\parallel}$ is a wavevector component along the periodic
trajectory, $L$ is the total orbit length, and $n$ is an integer
``quantum'' number.

For the $7442$ Hz mode, the primary orbit of interest is the
``bouncing ball" one.
The amplitude build-up along the
``bouncing ball" orbit in Fig. 1 (a) is clearly evident, and, by
inspection, we expect
the quantum number associated with this orbit to be $n = 7$.  Substituting
this
into Eq. (8) along with the appropriate $\alpha$ values for the orbit (i.e., 
$\alpha_1 = \alpha_2 = 0$), we obtain
$k_{\parallel} = 295$ m$^{-1}$.  Then, we may use Eq. (1) to estimate a 
frequency contribution of $f_{\parallel} = 6675$ Hz.  Of course, the total
frequency 
depends on $k_{\parallel}^{2} + k_{\perp}^{2}$, so we use the difference of
the 
observed $f$ and calculated $f_{\parallel}$ to estimate an effective
$k_{\perp} = 
100$ m$^{-1}$.  
This corresponds to an effective ``perpendicular" wavelength of 
$\lambda_{\perp} = 6.30$ cm,
which is comparable to the full width of the straight section of the plate.  
This observation is consistent with numerical studies of the Helmholtz 
equation, in which wavefunction scars of
``bouncing ball" trajectories have an effective perpendicular wavelength which
is slightly less than the full width of the straight section of the stadium 
boundary [1].

For the rectangular orbit that scars the $7622$ Hz mode, there are four
bounces
each with $\alpha_i = \pi/4$.  Using a quantum number of $n = 16$, we
determine that
$k_{\parallel} = 271$ m$^{-1}$ and $f_{\parallel} = 5644$ Hz, considerably
less than 
the measured frequency.  In this case, the
estimated perpendicular wavelength is rather small, $\lambda_{\perp} =
3.90$ cm.  
This can be understood with the following argument.
First, we note that the amplitude maxima of the rectangular orbit along the 
top straight edge of the plate
should occur very near to the classical trajectory which lies a distance 
$R(1-\frac{1}{\sqrt{2}}) = 1.17$ cm 
below the plate edge in Fig. 2.  Next, we consider the related problem of 
a long rectangular strip of plate which is 
clamped on one side and simply-supported on the other side (i.e., the amplitude
and its second normal derivative vanish).  Furthermore, the  width $l$ of
the strip 
is selected such that the amplitude maximum of the lowest
order vibrational mode occurs along the scarring orbit.  This 
problem is easily solved using standard methods [20, 22], and we find 
$l = 2.02$ cm.  We expect that the length $l$ should be close to the
above estimate based on the quantization condition 
for one-half of the perpendicular wavelength, $\lambda_{\perp}/2$. 
We find $\lambda_{\perp}/2 = 1.95$ cm, in good agreement.  

We have observed qualitatively similar scarring effects in clamped metal 
plates that have non-stadium boundary shapes, but which are predicted to be
chaotic
for the Helmholtz equation,
e.g., one-quarter of a ``bow-tie" shape [24].  We also
studied the spatial properties of vibrational modes in clamped plates with
a circular boundary; in this case no significant scarring was found.  
We attempted to measure spectral statistics, but the present experimental 
set-up is not suitable for obtaining accurate results.  The point
driving source does not allow detection of those modes with
nodal lines close to the driving point.  Furthermore, if one
moves the driving point to another location the eigenfrequencies 
shift due to coupling between the plate and the driving rod [8,24], 
so that one cannot combine spectra taken for different
drive positions in any analysis of spectral statistics. 

We have presented direct experimental evidence for scarring in the vibrational
modes of a thin metal plate in the shape of a stadium. We have also introduced
 a novel quantization condition
which is useful for
estimating the
experimental resonant frequencies of those modes that correspond to unstable
periodic orbits of the underlying classical billiard system.       
  It seems remarkable that
scarring effects - predicted on the basis of semiclassical theories 
that include no frictional
terms whatever - are clearly visible in this relatively low $Q$ system.   
Further, it is surprising that of the three wave patterns we have imaged 
two were scarred by short periodic orbits. The imaging was done only at those
excitation frequencies where the response of the plate was especially strong, since at the 
lower amplitudes the quality of interferograms is poor. At the 7442Hz and especially at 7622Hz 
excitation frequency one could actually hear the response of the plate. The Q-factors
at these excitation frequencies were also largest detected. It is possible that due to 
reduced overlap between scarred and neighboring non-scarred modes the leakage of energy 
is suppressed and the Q-factors are enhanced. If true, such an effect would be generic 
for wave systems exhibiting scarring, with relatively small,
but non-negligible losses. Further
experiments on bimorph quartz piezoelectric plates which would allow a better coupling 
scheme are being planned to explore this possibility 
and to improve on already existing measurements.

We thank Prof. K. Matveev for helpful discussions, and thank
Lester Chen, Dean Moyar, and Thomas Hund for help
with the experimental set-up.

\newpage

\begin{figure}
\caption{ (a) Holographic interferogram 
  for the $7442$ Hz mode. 
  The white dashed line superimposed on the hologram is a 
  guide to the eye. (b) Plate schematic. $\alpha$ and $\beta$ are the symmetry axis. $R_s$ is 
the position where the speaker needle was glued to the plate }
\end{figure}
 
\begin{figure}
\caption{ Holographic interferogram 
  for the $7622$ Hz mode.  
  A rectangular orbit is superimposed on the right half of the image
  as a guide to the eye.  }
\end{figure}

\begin{figure}
\caption{the proposed level diagram of an underlying pure system
unperturbed by the speaker-plate coupling and air damping. Each mode 
of an unperturbed plate is marked by the orbit that scars it. (-/+, -/+)
notation is used to show the symmetry class of a given mode. The first 
index shows the symmetry with respect to the $\alpha$ axis, the second with respect to the $\beta$
axis. Minus sign stands for odd symmetry, plus for even. }

\end{figure}

\end{document}